\documentstyle[twocolumn,epsfig,prl,aps]{revtex}

\begin{document}
\wideabs{

\title{Spinon excitation and M\"obius boundary condition in $S=1/2$ antiferromagnetic Heisenberg spin ladder with zigzag structure}
\author{Kouichi Okunishi and  Nobuya Maeshima$^1$}
\address{Department of Physics, Faculty of Science, Niigata University, Igarashi 2, Niigata 950-2181, JAPAN \\
$^1$Department of Physics, Graduate School of Science, Osaka University, Toyonaka, Osaka 560-0043, JAPAN \\
}
\date{\today}

\maketitle 

\begin{abstract}
We investigate the low-lying excitation of the antiferromagnetic zigzag spin ladder with M\"obius boundary condition.
Using the Lanczos and Householder diagonalization methods, we calculate the excitation spectrum of the zigzag ladder in the momentum space.
We then show that the topological defect generated by the M\"obius boundary provides the clear evidence of the single spinon excitation.
On the basis of the obtained single-spinon dispersion curve, we analyze the two-spinon scattering states under the usual periodic boundary condition.
We further discuss the interaction effect between the spinons, and the connection to the cusp singularity in the magnetization process.
\end{abstract}


}


Antiferromagnetic(AF)  $S=1/2$ Heisenberg  spin ladder with the zigzag structure   has been an important issue in theoretical studies of quasi one-dimensional(1D) quantum spin systems, since it is a minimal model including both of the quantum fluctuation and the frustrated interaction.\cite{mg,zigzag}
Particularly it is an interesting problem  to clarify how the effect of the frustration appears in the structure of the low-lying excitations,\cite{SS,mikeska} which is closely related to various characteristic behaviors of observable  physical quantities, such as the cusp singularity in the magnetization curve.\cite{cusp1,cusp2}

In analyzing the low-energy excitation of the zigzag antiferromagnet, an essential object is the spinon excitation, which may be described by a ``local-defect particle'' in the ground state.
However, the nature of the single spinon excitation can not been investigated directly for the system with the periodic boundary condition(PBC), which is assumed in the most of theoretical studies;
The total-$S$ of the system  under the PBC takes integer values  and hence one can not extract the single-spinon excitation of $S=1/2$ straightforwardly.

\begin{figure}
\begin{center}
\epsfig{file=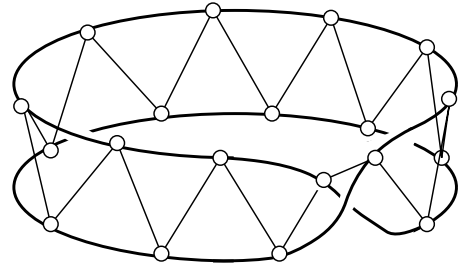,width=6cm}
\end{center}
\caption{The antiferromagnetic zigzag spin ladder with the M\"obius boundary.
The inter-ladder couplings, which are normalized to be unity, are denoted by thin line  and the intra-ladder couplings($\alpha$) are indicated by the thick line.}
\label{mobi}
\end{figure}

In order to discuss the topological-defect excitation, in this letter,  we deal with the AF zigzag spin ladder with the M\"obius boundary condition(MBC), which creates a domain-wall type defect in the system without losing the translational invariance.(See Fig.\ref{mobi})
In contrast to the usual PBC,  the zigzag ladder  with the M\"obios boundary contains the odd number of spins due to the network structure of the zigzag lattice, where the total-$S$ of the system can be labeled with a half integer.\cite{odd}
Then  an essential difference can be expected in the low-energy spectrum for the MBC, where the defect particle is able to move with a finite momentum.
We find that this defect particle, i.e. the M\"obius twist in the zigzag spin ladder, is identified as  the spinon excitation of $S= 1/2$.

In the following, we discuss the spinon excitation generated by the MBC, using the Lanczos and Householder diagonalization technique in the momentum space.
On the basis of the calculated spinon dispersion for the MBC, we further consider the low-energy excitation of the zigzag spin ladder with the PBC, where the two spinon scattering state plays an important role.
In addition, we show that the shape change of the spinon dispersion curve, which is connected with the cusp singularity in the magnetization process,
can be illustrated in the exact diagonalization spectrum obtained for the MBC.

The Hamiltonian of the zigzag spin ladder is given by 
\begin{equation}
{\cal H}=\sum_{i=0}^{N-1}\left[ \vec{S}_i\cdot\vec{S}_{i+1} + \alpha \vec{S}_i\cdot\vec{S}_{i+2} \right]  \label{zigzag}
\end{equation}
where $\vec{S}_i$ is the $S=1/2$ spin operator at $i$-th site and $\alpha$ is the ratio of the inter- and intra-ladder couplings.
The number of sites is denoted as $N$, which takes odd integer for the MBC and even number for the PBC. 
Here we note that, for the  infinite-length system,  the ground state of Eq. (\ref{zigzag}) for $\alpha < \alpha_c\equiv 0.2411$ is in the spin liquid phase and that for $\alpha > \alpha_c $ is in the dimer gapped phase.\cite{nomura} 

An important feature of the M\"obius boundary is that the translational invariance of the system is maintained, although the domain wall is inserted in the system.
This implies that the topological defect can move around in the ``M\"obius ribbon'' smoothly.
(In the Fig. \ref{mobi}, the twisted part can move  in the whole region of the ``M\"obius ribbon''.) 
Thus we can make the Hamiltonian (\ref{zigzag}) block-diagonal with respect to  the momentum quantum number $k_l=2\pi l/N$, where $l=0,1, \cdots, N-1$.
We perform the Lanczos diagonalization up to 28 sites and the Householder diagonalization up to 18 sites in the momentum space.

Before proceeding to a discussion of the computed results, here,  it should be noted that, in the spin liquid phase, there are gapless excitations near the ground state, which requires us to  analyze the calculated data precisely.
Thus, in this paper, we consider the dimer-gapped phase  $\alpha_c \le \alpha \le 1$ mainly, although the argument of the MBC is also available for the spin liquid phase.
The single-spinon excitation in the dimer phase has been discussed only near the Majumdar-Ghosh point($\alpha=1/2$), using the variational approach combined with the matrix-product ground-state.\cite{SS,mikeska}
In the context of the spin-Peierls system, the dynamical properties of the zigzag spin ladder is also investigated with a variety of the numerical techniques.\cite{sorensen,watanabe}
However, the topological boundary condition has not been emphasized so far.
In the following,  the striking evidence of the single spinon excitation is demonstrated in the exact diagonalization spectrum of the system with the MBC.

\begin{figure}
\begin{center}
\epsfig{file=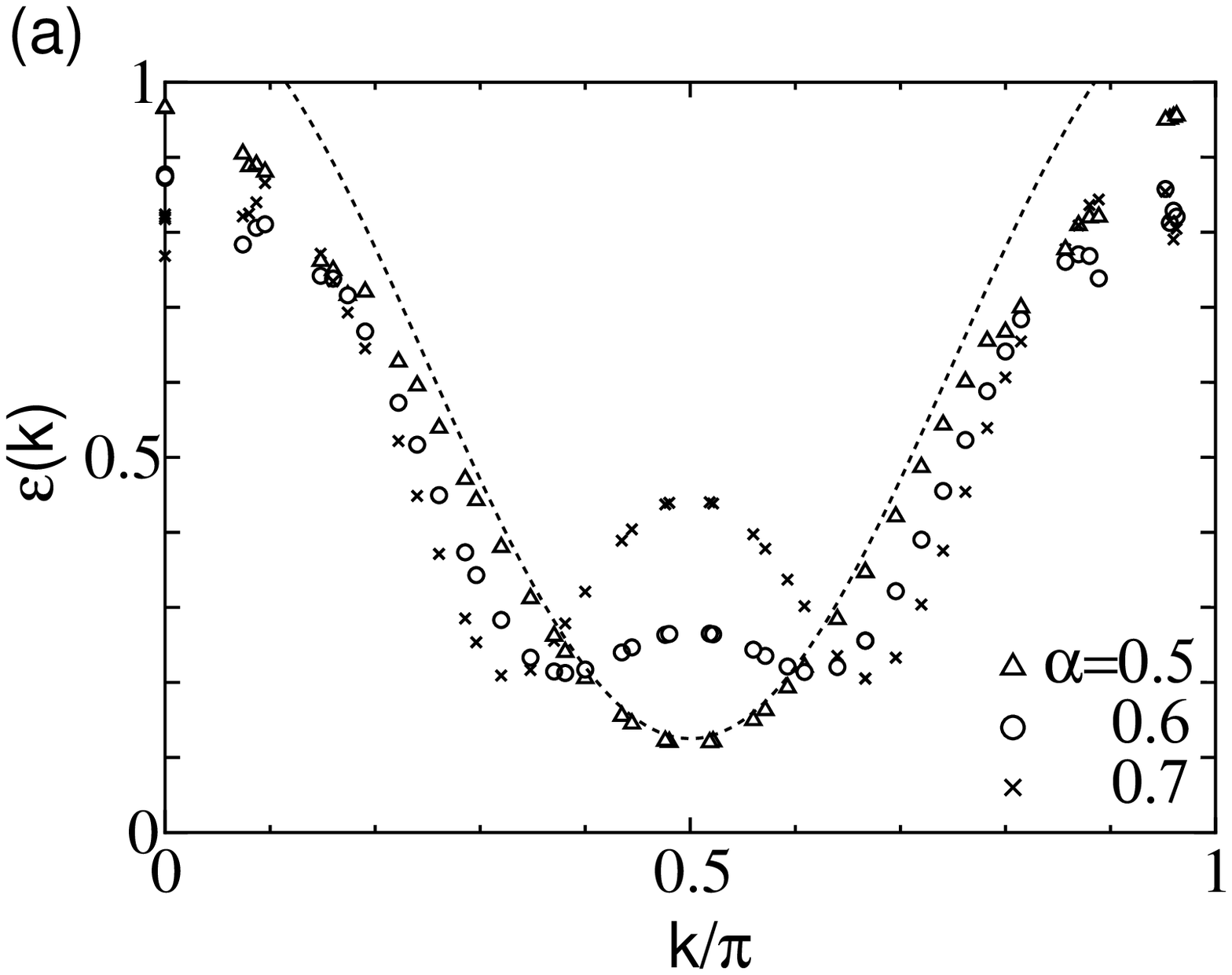,width=7cm}\\

\epsfig{file=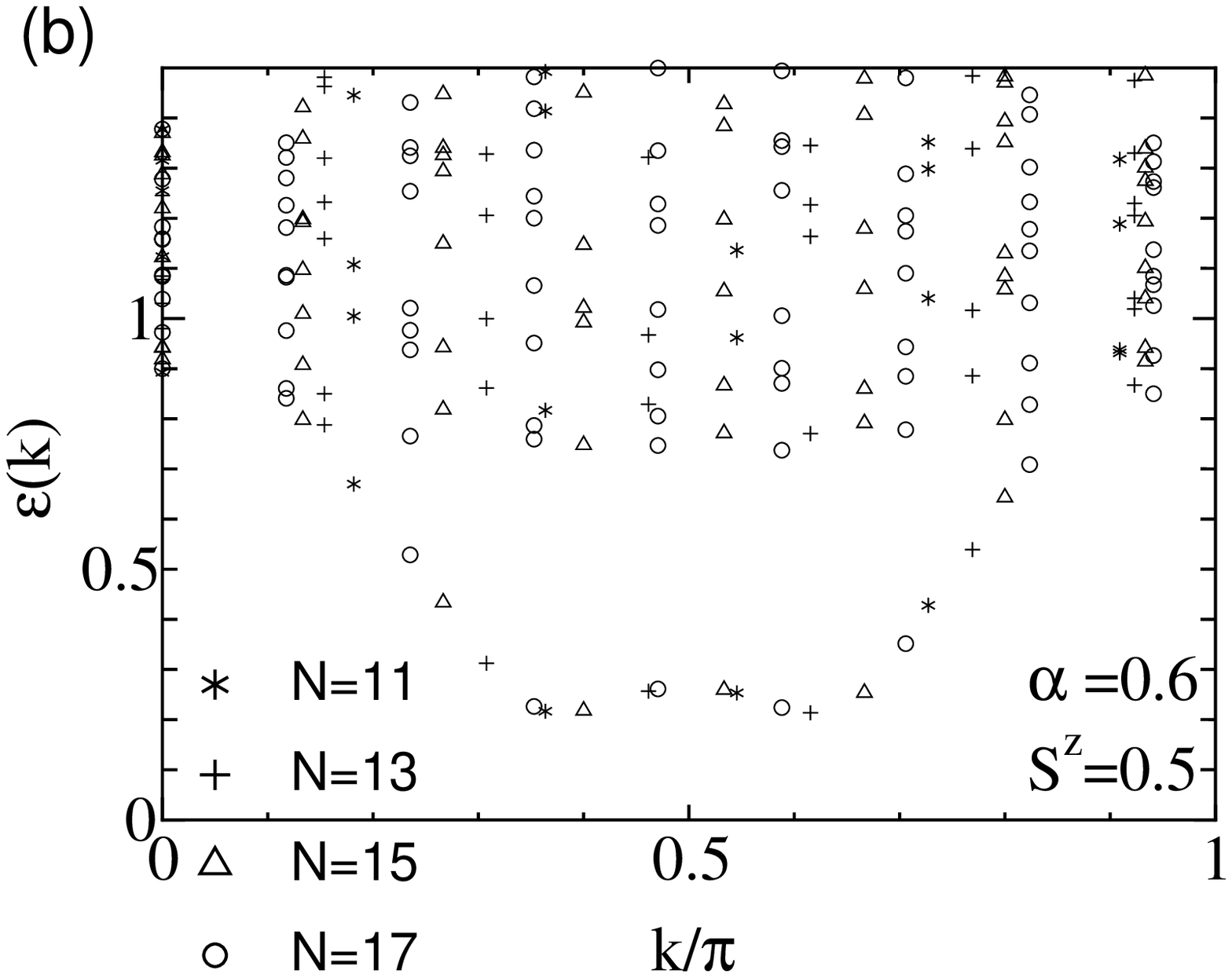,width=7cm}
\end{center}
\caption{The one-spinon dispersion curve of the zigzag chain with the M\"obius boundary condition. (a) Lanczos results up to 27 sites for $\alpha=0.5$(Majumdar-Ghosh point), $0.6$ and $0.7$. The dashed line is the spinon dispersion curve obtained with the variational calculation for $\alpha=0.5$. (b) The spectrum obtained with Householder diagonalization for $\alpha=0.6$} \label{lncodd}
\end{figure}

In Fig. \ref{lncodd}-(a), we show the low-energy spectrum of the zigzag spin ladder of $\alpha=0.5$, $0.6$ and $0.7$, which are obtained with the Lanczos diagonalization up to 27 spins in the total-$S^z=1/2$ subspace.
We also show the Householder diagonalization results for $\alpha=0.6$ in  Fig. \ref{lncodd}-(b), which include the higher-energy spectrum.
Since  the defect particle of the MBC can move in the ring, the low-energy spectrum is described with  the continuous curve, which is  adiabatically connected to the ground state of the MBC system.
In other word, the ground-state energy for the MBC is given by the bottom of the dispersion curve of the defect particle, which is located at near $k=\pi/2$.
In Fig. \ref{lncodd}, however,  we have determined the origin of the energy axis with  
\begin{equation}
E^{\rm MBC}_0(N)=e_g \times N,
\end{equation}
where $e_g$ denotes the ground-state energy per site obtained from the ground-state energy for the PBC system of 28 sites with use of  $e_g=E_0^{\rm PBC}(N)/N$. 
For example, we have $e_g=-0.375$(exact), $-0.3802$, and $-0.3972$ for $\alpha=0.5$, $0.6$ and $0.7$ respectively, where the finite-size correction can be negligible.
Then,  we find that the computed spectra for the various $N$ get on a single curve, which can be regarded as the one-particle dispersion curve  above the ``virtual ground state'' with the ``virtual excitation gap''.\cite{shift}
Moreover, in Fig. \ref{lncodd}-(b), the dispersion curve of the defect particle is clearly separated below the continuum spectrum, which implies that the particle is a well defined object in the zigzag spin ladder.
Thus it will be allowed for us to call this defect particle in the MBC system as a ``spinon excitation''. 
In fact, the dispersion curve for $\alpha=0.5$ is in good agreement with  the spinon dispersion obtained with the variational calculation\cite{SS}, which is represented as the dashed line in the Fig. \ref{lncodd}-(a).

\begin{figure}
\begin{center}
\epsfig{file=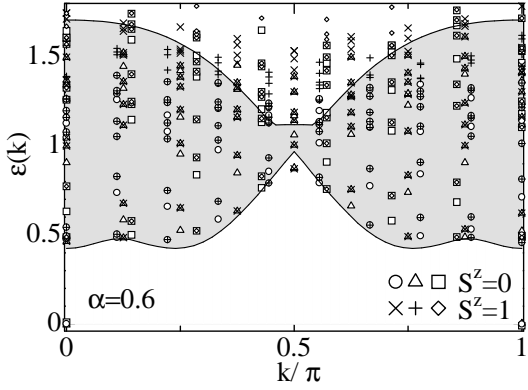,width=7cm}
\end{center}
\caption{Hoeseholder diagonalization results for $\alpha=0.6$ with the periodic boundary condition. The shaded region represents the two-spinon continuum.} \label{hh06}
\end{figure}

In Fig. \ref{lncodd}, we can also see  that the shape of the dispersion curve changes into the double well structure,  which is expected from the cusp singularity in the magnetization process.
Such a clear evidence of the shape change is difficult to find for the PBC case, where the two-spinon continuum  should be dealt with numerically.(See also Fig.\ref{hh06})
Precise calculations indicate that this shape change occurs at $\alpha_{\rm cusp}\simeq 0.54$, which is slightly larger value than  the Majumdar-Ghosh point $\alpha=1/2$.
We note that the similar behavior of the dispersion curve is reported in the $S=1$ bilinear-biquadratic spin chain,\cite{blbq,disp}  which exhibits the magnetization cusp\cite{cusp1} and the exact matrix product state called valence-bond-solid.\cite{AKLT}

Let us next investigate the  excitation spectrum of the PBC case, exploiting the spinon dispersion curve extracted from the MBC.
In Fig. \ref{hh06},  we display the energy spectrum of the PBC for  $\alpha=0.6$, which is computed with Householder diagonalization method   up to 18 sites.
For the case of the PBC, the dimerized ground-state is located at $k=0$ and $\pi$, and there is the  excitation gap whose magnitude is twice of the virtual gap for the MBC.
This magnitude of the gap is consistent with the result of the density matrix renormalization group.\cite{white}
Further we can see that the singlet and triplet excitations conform the continuum spectrum above the excitation gap, which can be represented as the scattering state of the two spinons.
(In Fig.1, this will be illustrated as the two twists running in the zigzag ribbon independently.)

To see this,  we compare the free two-spinon continuum with the Householder results.
The two-spinon continuum is given by
\begin{eqnarray}
E(q)&=&\varepsilon(k_+)+\varepsilon(k_-),\\
q&=&k_++k_-,
\end{eqnarray}
where $\varepsilon(k)$ is the single-spinon dispersion curve, which can be determined from the result of the  MBC system.
We find that  $\varepsilon(k)$ is  well fitted with the form:
\begin{eqnarray}
\varepsilon(k)=\sqrt{A(k)B(k)}, \label{fitting}
\end{eqnarray}
where $A(k)= 1+a_1\cos(2k)+a_2\cos(4k)$ and  $B(k) = b_1+b_2\cos(2k)+b_3\cos(4k)$ with the fitting parameters $\{a\}$ and $\{b\}$.\cite{fitting}
The calculated two-spinon continuum is shown  as the shaded region in Fig. \ref{hh06}.
Clearly the lower bound of the two-spinon continuum can explain the diagonalization results, including the subtle structure above the excitation gap near $k=0$ and $\pi$.
Hence we can conclude that the excited state in the PBC system is basically described with the two-spinon scattering state.

As was seen in the above, the low-energy  property of the PBC system is well  explained with the free two-spinon state. 
For the purpose of the quantitative analysis, however,  the interaction effect between the spinons becomes important, especially in the finite-size system.
In addition,  we can see that the eigenvalues in the  $S^z=1$  subspaces deviate from those of $S^z=0$ even in the low-energy region(see Fig. \ref{hh06}).
This is because the frustrated interaction  affects the excitation spectrum.
Thus, let us next investigate the spinon-spinon interaction effect on the low-energy spectrum.
For this purpose, we consider the dilute spinon gas model, which may be described with  an effective  Hamiltonian with two-body interaction:
\begin{eqnarray}
&{}&{\cal H}_{\rm eff}=\sum_{k,s} \varepsilon(k) a^\dagger_{k,s} a_{k,s}\nonumber \\
 &{}& +  \sum_{k,k',s,s'} V_{s,s'}(k,k')  a^\dagger_{k,s} a^\dagger_{k,s} a_{k',s'} a_{k',s'},\label{effectivebg}
\end{eqnarray} 
where $a^\dagger_{s,k}(a_{s,k})$ is the spinon creation(annihilation) operator and $V_{s,s'}(k,k')$ is the effective coupling.
The index  $s$ denotes the $z$-component of the spinon excitation and the single-spinon dispersion curve $\varepsilon(k)$ has been determined with Eq. (\ref{fitting}).

In order to discuss the interaction effect in the exact diagonalization spectrum, let us now concentrate on the two-spinon problem, for which the Schr\"odinger equation can be written down explicitly:
\begin{eqnarray}
\left[\varepsilon(k_+)+\varepsilon(k_-)-E(q)\right]\psi_s(k_+,k_-)= \nonumber \\
-\sum_{k',s'}V_{s,s'}(k,k')\psi_{s'}(k'_+,k'_-), \label{2body}
\end{eqnarray}
where $k_{\pm}= q/2 \pm k$ and  $q$ is the momentum of the center of mass.
The spin symmetry can be classified in the spin singlet and triplet sectors.
Assuming the repulsive  $\delta$-function-type interaction $V_{s,s'}(k,k')=c$(=const), the solution of (\ref{2body}) is given by
\begin{equation}
1=c\sum_k\frac{ 1}{E(q)-\varepsilon(k_+)-\varepsilon(k_-)},\label{2bodyans}
\end{equation}
We solve Eq. (\ref{2bodyans}) for a finite-size system numerically and compare the results  with the diagonalization ones.
Particularly we consider the system of the $N=28$, where the wave number is given by $k_l=2\pi l/N$ with $l=0,1, \cdots, 27$.

\begin{figure}
\epsfig{file=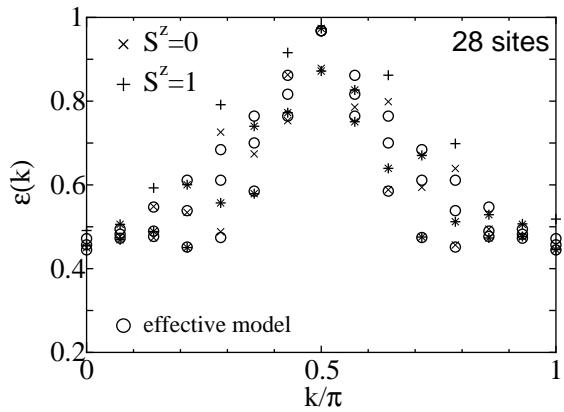,width=7.5cm}
\caption{Comparison of the Lanczos results with the effective spinon gas model. }
\label{effspect}
\end{figure}

We first discuss the spectrum obtained with the  Lanczos method in the total-$S^z=0$  subspace, where the singlet state and the $S^z=0$ state of the triplet appear.
Fig. \ref{effspect}  shows the results for the two-body problem  (\ref{2body}) with $c=0.1$, where we can see the quantitative agreement in the Lanczos results of $S^z=0$ sector.
If we drop the interaction term(i.e. free-spinons case), the second lowest and the  third lowest eigenvalues of the effective model degenerate and then a branch of the spectrum  can not be explained. 
Thus we can conclude that the Lanczos spectrum of total-$S^z=0$ is well explained by the effective spinon gas model with the repulsive interaction.

We next proceed to the analysis of the $S^z=1$ sector.
In Fig. \ref{effspect}, we can see that the Lanczos results of the triplet sector are rather scattered, comparing with those of the singlet one.
Such a difference between the singlet and triplet spectrum in the low-energy region may be a characteristic behavior in the strongly frustrated system, in contrast to  the Heisenberg spin chain where the triplet and singlet excitations exhibit a quite good correspondence.
In fact, we find that this kind of singlet-triplet difference emerges, as $\alpha$ is increased beyond $\alpha\simeq 0.4$.
Accordingly,  some of the spectrum of the triplet sector do not agree with the present analysis for $\alpha=0.6$ utilizing the simple $\delta$-function interaction;
The effect of the frustration may appear in the triplet excitations significantly, and the interaction term $V_{s,s'}(k,k')$ can be expected to have more complicated  form than that of the singlet.
Here, we remark that the finite-size effect in the triplet sector is much bigger than the singlet sector, which is consistent with the observation of the strongly affected triplet sector.
In order to verify the dilute spinon gas model for the triplet, however, it is required to investigate the excitation spectrum of a longer system.

Finally we  make a comment on the bound state.
In the variational treatment of the two spinon excitation,  $V_{s,s'}(k,k')$ is not a simple repulsive interaction, where the bound state of the two spinon can appear near $k=\pi/2$.\cite{SS,mikeska,sorensen} 
In the diagonalization results,  we can see that there are lower energy eigenvalues at $k=\pi/2$ than the scattering state, which is consistent with  the bound state.

To summarize, we have discussed the M\"obius boundary condition(MBC) in the antiferromagnetic Heisenberg spin ladder with the  zigzag structure.
The MBC introduces the topological defect that is described  with the domain-wall type particle of $S=1/2$.
Using the exact diagonalization method in the momentum space, we have found that this defect particle gives the clear evidence of the spinon excitation.
In addition, we have shown  the shape change of spinon dispersion curve, which is associated with the cusp singularity of the magnetization process.

We have investigated  the low-energy spectrum with the  usual periodic boundary condition(PBC), where the two spinon state plays an important role.
We have shown that the two spinon scattering state explains well the spectrum computed with the exact diagonalization.
We have further discussed the interaction effect between the spinons, using the dilute spinon gas model.
The $S^z=0$ sector is in good agreement with the effective model, while the  spectrum of $S^z=1$ subspace exhibits rather complicated behavior.
To clarify the role of the frustrated interaction in the triplet sector, a further analysis of the zigzag system will be required.

The authors would like to thank T. Nishino for valuable comments.
A part of the present numerical computations was performed at Yukawa Institute Computer Facility.



\begin{thebibliography}{99}
\bibitem{odd} A system with odd number of spins has been considered in Ref.\cite{sorensen}. However the topological nature of the boundary condition of the system has not been mentioned.
\bibitem{shift}If we take another value of $e_g$, the spectrum is scattered unlike in Fig. \ref{lncodd}
\end{thebibliography}
\end{document}